# A Toy Model of Renormalization and Reformulation

Vladimir Kalitvianski[1*]

[1]Direction d'Energie Nucléaire, Commissariat à l'Energie Nucléaire, Grenoble, France
*Corresponding author: vladimir.kalitvianski@numericable.fr

**Abstract** I consider a specially designed simple mechanical problem where a "particle acceleration" due to an external force creates sound waves. Theoretical description of this phenomenon should provide the total energy conservation. To introduce small "radiative losses" into the phenomenological "mechanical" equation, I advance first an "interaction Lagrangian" similar to that of the Classical Electrodynamics (kind of a self-action ansatz). New, "better-coupled" "mechanical" and "wave" equations manifest unexpectedly wrong dynamics due to changes of their coefficients (masses, coupling constant); thus this ansatz fails. I show how we make a mathematical error with advancing a self-interaction Lagrangian. I show, however, that renormalization of the fundamental constants in the wrong equations works: the original "inertial" properties of solutions are restored. The exactly renormalized equations contain only physical fundamental constants, describe well the experimental data, and reveal a deeper physics – that of permanently coupled constituents. The perturbation theory is then just a routine calculation only giving small corrections. I demonstrate that renormalization is just illegitimately discarding harmful corrections fortunately compensating this error, that the exactly renormalized equations may sometimes accidentally coincide with the correct equations, and that the right theoretical formulation of permanently coupled constituents can be fulfilled directly, if realized.

## 1. Introduction

Any macroscopic body has interacting "constituents", internal degrees of freedom, collective modes of relative and global motion, and we observe this body with help of exchange of energy with these collective modes. This exchange is essentially already encoded in the laws of motion of our macroscopic body by determining the properties of our phenomenological equations.

If we "push" one of constituents, we not only transfer energy-momentum to the body as a whole, but also excite internal degrees of freedom, the latter excitation being inelastic losses. Sometimes the inelastic losses are small compared to the global energy exchanged and we are inclined to forget about them and think of the body as of an "elementary" one. In particular, such an "elementary particle" like electron is sincerely thought of as a free one despite permanent interaction with "its own electromagnetic degrees of freedom", the latter being thought to be somewhat "independent" of electron. That is why when we are trying to "switch on" the interaction "once again", we get conceptual and mathematical problems. Unfortunately, these problems are currently understood as Nature properties rather than as our errors.

In the literature there are many papers with toy models and many books – all explaining usefulness and "naturalness" of renormalization and renormalization group in QFT. The present article is, on the contrary, a word in favor of a *reformulation approach* that in former times was a mainstream activity and is abandoned today.

Here I would like to sketch out how and when we fall in conceptual and mathematical error while advancing our theories. Namely, I explain the true reason of difficulties encountered in course of coupling equations and the meaning of renormalization of the fundamental constants. Briefly speaking, the problems occur because our understanding of physics and our way of coupling are wrong. At the same time, we may be very close to a better understanding and we may keep practically the same old equations in a correct formulation. My consideration is on purpose carried out on a simple and feasible mechanical problem in order to demonstrate unambiguously that the fundamental constant modifications are not quantum, or relativistic, or non-linear "physical effects" occurring with "bare" particles, but errors admitted in our equation guessing. Classical Electrodynamics (CED) and Quantum Electrodynamics (QED) and their historical developments will be used as examples to follow in our Classical Mechanics problem.

In Section 2 I outline the experimental setup and the corresponding phenomenological equations. These equations are analogous to the CED equations without radiation reaction force. In Section 3 I advance an "interaction Lagrangian" in order to derive the radiation reaction force necessary for obtaining the energy conservation law. I show that despite achieving formally an "energy conservation law", the new coupled equations differ from the original ones not only with the presence of a radiation reaction force, but also with other terms that essentially modify the dynamics of our variables. Thanks to a specially designed (mechanical one-mode) problem, the origin of coefficient modifications is clearly visible whereas in QED it is obscured due to the perturbative rather than exact treatment of the "interaction Lagrangian". In Section 4 I fulfil renormalization of the modified constants and arrive at good equations resembling the old ones and containing solely correct radiation reaction force, as was planned in the beginning of Section 3. I analyze the physics contained in these exactly renormalized equations and show that it finally corresponds to our particular mechanical problem. In Section 5 I give a general discussion of the presented material and its further implications.

## 2. Phenomenon to describe

The difficulties encountered in CED and QED can be modeled quite reasonably with Classical Mechanics. It is instructive to analyze them to the end. For that, let us consider a macroscopic probe body constructed with a purpose to model a one-mode compound system (Fig. 1).

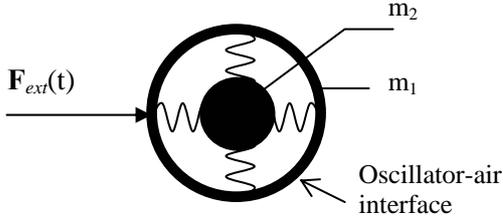

**Fig. 1**. Mechanical model of a one-mode compound system

We will suppose that it is a rather rigid shell of a diameter $D$ with a core inside connected with springs modelling a 3D oscillator. From the exterior the probe body looks as a solid ball of a mass $M_p = m_1 + m_2$ and experimentalists do not know its true composition. Neither do we theorists. The energy of oscillations is assumed to be small compared to the body kinetic energy, and experimentally one first establishes Newton equation for this body motion (hereafter also called a "particle"):

$$M_p \ddot{\mathbf{r}}_p = \mathbf{F}_{ext}(\mathbf{r}_p \mid t). \qquad (1)$$

Eq. (1) contains such fundamental physical notions as a particle position $\mathbf{r}_p$, a particle mass $M_p$, and an external force $\mathbf{F}_{ext}$, all of them being measurable physical quantities and nothing is "bare". As our particle is not really point-like, Eq. (1) is written, strictly speaking, for its geometric center position. Eq. (1) is our analogue to the Lorentz equation for a point-like charge $q$ in an external electromagnetic field without radiation reaction force [1]:

$$m_q \ddot{\mathbf{r}}_q = q\sqrt{1 - \frac{\dot{\mathbf{r}}_q^2}{c^2}} \left[ \mathbf{E}_{ext} + \frac{1}{c}\dot{\mathbf{r}}_q \times \mathbf{B}_{ext} - \frac{1}{c^2}\dot{\mathbf{r}}_q \left( \dot{\mathbf{r}}_q \cdot \mathbf{E}_{ext} \right) \right]. \quad (2)$$

The "particle" (1) Lagrangian in an external time-dependent potential is

$$L_p = \frac{M_p \dot{\mathbf{r}}_p^2}{2} - V_{ext}\left(\mathbf{r}_p \mid t\right). \qquad (3)$$

It is similar to a charge Lagrangian with a known external electromagnetic potential $A_{ext}$:

$$L_q = -m_q c^2 \sqrt{1 - \frac{\dot{\mathbf{r}}_q^2}{c^2}} - q\varphi_{ext} + \frac{q}{c}\dot{\mathbf{r}}_q \cdot \mathbf{A}_{ext}. \qquad (4)$$

Now let us suppose that after our getting well accustomed to describing our "particle" with (1), experimentalists discovered that even the simplest *acceleration* of our probe body by a constant external force created some weak sound waves with a characteristic frequency $\omega$. We will suppose that the waves were discovered experimentally (rather than predicted) and some phenomenological description was then established. For example, experimentalists varied the force absolute value $F_{ext}$ and its duration $T$ and found an empiric equation for the oscillation amplitude. Finally, they found that the observed sound amplitude $A_{sound}$ at some distance $S \gg D$ is a solution to the following driven oscillator equations (sound damping neglected for instance):

$$\ddot{A}_{sound} + \omega^2 A_{sound} = \alpha_{sound}(S)\, \mathbf{n} \cdot \ddot{\mathbf{r}}_p(t). \qquad (5)$$

This equation is mathematically quite analogous to a CED equation for the amplitude $\mathbf{E}_k(t)$ of a Fourier harmonic of the transverse electric field (a standing wave) "sourced" with the charge acceleration $\ddot{\mathbf{r}}_q(t)$. (Namely the CED equation for one harmonic $\mathbf{E}_k(t)$ has inspired me to design this mechanical toy model - all other harmonics are excited in the same way so we can study *one* for simplicity). Here $\alpha_{sound}(S)$ is an experimentally measurable dimensionless coefficient of efficiency of wave excitation with the particle acceleration (a "coupling" or "pumping efficiency" constant) and the unit vector $\mathbf{n}$ points out the sound propagation direction from the body. The body velocity $\dot{\mathbf{r}}_p$ (Doppler effect) and the distance $S$ (sound retardation) are assumed reasonably small and thus inessential in our model.

The sound wave amplitude $A_{sound}$ at a long distance $S \gg D$ from our "particle" depends, of course, in a known way on this distance and on the angle $\theta$ between the force and the propagation direction, and we will assume that the measured sound amplitude is proportional to some "true oscillator" amplitude $\mathbf{r}_{osc}$, although we do not know much about the true oscillator yet, due to, say, imperfections of experimental facility (for example, it is large compared to $D$) and thus difficulties in precise local ($S \approx D$) measurements. In the following we will suppose that Eq. (5) can be rewritten factually for the 3D oscillation amplitude $\mathbf{r}_{osc}$ defined from the proportionality $A_{sound} \propto \mathbf{n} \cdot \mathbf{r}_{osc}$ (kind of limit $S \to D/2$ in Eq. (5)). Then the oscillator Eq. (5) can be equivalently written via $\mathbf{r}_{osc}$, oscillator (unknown) mass $M_{osc}$, and (unknown) spring constant $k$, i.e., in a more canonical way:

$$M_{osc}\ddot{\mathbf{r}}_{osc} + k\mathbf{r}_{osc} = \alpha M_{osc}\ddot{\mathbf{r}}_p(t), \quad \omega = \sqrt{k/M_{osc}}. \quad (6)$$

As the measured body rigidity, mass $M_p$, and the size $D$ do not provide the observed and unique frequency, our experimentators continue performing their experiments, and we theorists get busy with the existing description of this phenomenon (1), (6). In particular, it is natural to assume that it is the true oscillator energy that is spent on creating sound waves, the latter being the reason of oscillation damping.

The coupling constant $\alpha$ of the true oscillator does not depend on any distance by definition (as if it were something like $\alpha = \alpha(S \approx D/2)$ ). The oscillator Lagrangian corresponding to (6) is then the following:

$$L_{osc} = \frac{M_{osc}\dot{\mathbf{r}}_{osc}^2}{2} - k\frac{\mathbf{r}_{osc}^2}{2} - \alpha M_{osc}\dot{\mathbf{r}}_{osc}\dot{\mathbf{r}}_p(t), \quad (7a)$$

where $\dot{\mathbf{r}}_p(t)$ is a known (given) function of time, namely, the solution to (1). The right-hand side (driving or "pumping") term $\alpha M_{osc}\ddot{\mathbf{r}}_p(t)$ in (6) is then obtained from the kinetic part of Lagrange equation $\frac{d}{dt}\frac{\partial L_{osc}}{\partial \dot{\mathbf{r}}_{osc}}$. Eqs. (1) and (6) are our starting point for guessing a "better-coupled" system of equations.

Eq. (6) can, however, be equivalently rewritten via the external (known) force:

$$M_{osc}\ddot{\mathbf{r}}_{osc} + k\mathbf{r}_{osc} = \alpha\frac{M_{osc}}{M_p}\mathbf{F}_{ext}(\mathbf{r}_p(t)). \quad (8)$$

The latter unambiguously shows that the point of external force application (particle) permanently belongs to the oscillator. As well, Lagrangian for (8) can be different in form from (7a) (the difference is a full time derivative):

$$L_{osc} = \frac{M_{osc}\dot{\mathbf{r}}_{osc}^2}{2} - k\frac{\mathbf{r}_{osc}^2}{2} + \alpha\frac{M_{osc}}{M_p}\mathbf{r}_{osc} \cdot \mathbf{F}_{ext}(\mathbf{r}_p(t)), \quad (7b)$$

i.e., *the starting point form is not unique*. These facts will be "rediscovered" later on, after fulfilling the exact renormalization (see formula (26)), and now we will close our eyes on it and will base our theoretical developments on (6) and (7a) where the pumping term is given via the particle dynamical variables $\dot{\mathbf{r}}_p(t)$ and $\ddot{\mathbf{r}}_p(t)$, like in CED/QED.

Let us note that as soon as the external force stops acting, the "wave system" (6) or (8) decouples from the "mechanical" one (1) and becomes "free". This makes an impression that the "mechanical" and the "wave" systems only interact during body acceleration phase, otherwise they are "independent". This (false) impression will also strongly impact our further theoretical development.

So, at some stage we may well have two equations (1) and (6) established experimentally with help of macroscopic measuring devices describing some non trivial physical phenomena occurring with our macroscopic body, like it was the case in Electrodynamics. Next comes our purely theoretical reasoning. For example, although we do not know the body composition, we may sincerely think it is simple, point-like so that three degrees of freedom suffice to describe it. It is the same what we usually think of the electron in Electrodynamics – a point-like elementary object. Now we will try to make ends meet within our simplified "point-like" model.

The main theoretical concern is that we think the theory is not fully developed yet – the particle energy possible variations during acceleration are not related to the oscillator energy gain: $dE_p = \frac{\partial V_{ext}}{\partial t}dt \neq -dE_{osc}$ $= -\alpha M_{osc}\dot{\mathbf{r}}_{osc}\ddot{\mathbf{r}}_p dt$. It looks like while establishing experimentally Newton Eq. (1) the corresponding "radiative losses" were not noticeable due to their smallness: $|\Delta E_{osc}| << |\Delta E_p|$.

## 3. Theory Development I: Better-Coupled Equations

Now we theorists may want to intervene in order to "reestablish" the energy conservation law. In CED such an intervention led to a mass addendum and to the need of mass renormalization, and here we will follow the CED logic. Namely, we suppose that Eq. (1) is not exact and needs a "radiation resistance" force; Eq. (6) and its solutions being still practically perfect as experimentally justified. Thus, our intention and goal is to preserve solutions of Eq. (6) (that we understand as preserving the form of Eq. (6) as it is) and add a "radiation reaction" term into the particle Eq. (1) so that the total energy gets conserved.

When $\alpha = 0$, Eqs. (1) and (6) are decoupled and have their own independent Lagrangians, denoted as $L_p^{(0)} = \frac{M_p\dot{\mathbf{r}}_p^2}{2} - V_{ext}(\mathbf{r}_p)$, $L_{osc}^{(0)} = \frac{M_{osc}\dot{\mathbf{r}}_{osc}^2}{2} - k\frac{\mathbf{r}_{osc}^2}{2}$. In order to derive the searched "radiation reaction" term for the particle equation, let us construct an "interaction Lagrangian" $L_{int}$ to them, i.e., the total system Lagrangian, denoted hereafter as $L_{Trial}$, is a sum of all of them: $L_{Trial} = L_p^{(0)} + L_{osc}^{(0)} + L_{int}$. Let us try this one:

$$L_{int} = -\alpha M_{osc}\left(\dot{\mathbf{r}}_p\dot{\mathbf{r}}_{osc} - \frac{\eta}{2}\dot{\mathbf{r}}_p^2\right). \quad (9)$$

Here the cross term $\propto \dot{\mathbf{r}}_p\dot{\mathbf{r}}_{osc}$ gives the desired pumping term $\propto \ddot{\mathbf{r}}_p$ in the wave equation from $\frac{d}{dt}\frac{\partial L_{int}}{\partial \mathbf{v}_{osc}}$. Its *form* is evidently borrowed from $L_{osc}$ (7a) mentioned above (i.e., it is written *by analogy* with (7a)), but now we think that both $\dot{\mathbf{r}}_p$ and $\dot{\mathbf{r}}_{osc}$ must be *unknown* variables in it. Considering $\dot{\mathbf{r}}_p$ an unknown variable in (9) is our **ansatz** that will make a difference between "insufficiently coupled" Eqs. (1), (6) and the new ones being derived. We think it will simply add the necessary radiation reaction force in Eq. (1).

The quadratic term $\propto \eta \dot{\mathbf{r}}_p^2$ in (9) simulates here a self-action contribution analogous to the CED/QED electromagnetic mass and is to some extent a stretch in our mechanical problem (it may correspond to the self-action contribution of the short-range force experienced with other bodies colliding with our "particle"), but we may always advance it with saying that it does not enter into the oscillator equation, so Lagrangian (9) is the "most general one satisfying our requirements".

Lagrangian (9) may be called a "self-interaction" Lagrangian similar to that of CED/QED where both the current $j_\mu$ and its field $A_\mu$ are considered unknown and coupled variables, and the corresponding Lagrangian $L_{int} = -q\varphi + \dfrac{q}{c}\dot{\mathbf{r}}_q \cdot \mathbf{A}$ is added to (4). The difference with our mechanical model (1), (6), (9) is mainly in keeping in (9) only one oscillator and in assuming finiteness of the "self-energy" contribution $\propto \eta$ because we do not need an infinite $\eta$ for our modest purposes.

New, "better-coupled" mechanical and wave equations are the following:

$$\begin{cases} M_p \ddot{\mathbf{r}}_p = \mathbf{F}_{ext}(\mathbf{r}_p \mid t) + \alpha M_{osc}\left(\ddot{\mathbf{r}}_{osc} - \eta \ddot{\mathbf{r}}_p\right), \\ M_{osc}\ddot{\mathbf{r}}_{osc} + k\,\mathbf{r}_{osc} = \alpha M_{osc}\ddot{\mathbf{r}}_p. \end{cases} \quad (10)$$

At first glance the oscillator equation has not changed and the particle equation has acquired some "radiation reaction" terms, as we planned.

As well, the zeroth-order approximation of Eqs. (10) ($\alpha = 0$) corresponds to an "elastic" particle behavior – whatever the external force is, no oscillations are excited. Thus, our system behavior in this approximation is similar to that of QED – the textbooks have plenty of such "elastic" scattering results: the Mott (Rutherford), Bhabha, Klein-Nishina cross sections, etc.

In the first perturbative order we obtain "radiative corrections" and creation of waves of frequency $\omega$ due to the particle acceleration $\ddot{\mathbf{r}}_p^{(0)}(t)$ (à la Bremsstrahlung), etc. In other words, at first (perturbative) glance the "better-coupled" equation system may look acceptable.

### 3.1. Energy conservation in Eqs. (10)

With Noether theorem or directly from Eqs. (10) we obtain the following conservation law (case $\partial V_{ext}/\partial t = 0$):

$$\frac{d}{dt}\left(E_p + E_{osc} + L_{int}\right) = 0. \quad (11)$$

The quantity $E_p + E_{osc} + L_{int}$ is thus conserved. (Here we have a plus sign at $L_{int}$ because $L_{int}$ is of a "kinetic" rather than of "potential" nature.)

The energy conservation law (11) does not look as $E_p + E_{osc} = const$. There is an additional term here. However, it is quite similar to the CED conservation law with the "resistance" force $\propto \dddot{\mathbf{r}}_q$ where an extra term is also present in the power balance. Compare the CED conservation law (non relativistic approximation):

$$\frac{d}{dt}\left(E_q + \delta m_q \frac{\dot{\mathbf{r}}_q^2}{2}\right) = -\left(\frac{2e^2}{3c^3}\ddot{\mathbf{r}}_q^2\right) + \frac{2e^2}{3c^3}\frac{d}{dt}(\dot{\mathbf{r}}_q\ddot{\mathbf{r}}_q), \quad (12)$$

with ours:

$$\frac{d}{dt}\left(E_p + \alpha M_{osc}\eta\frac{\dot{\mathbf{r}}_p^2}{2}\right) = -\left(\frac{d}{dt}E_{osc}\right) + \alpha M_{osc}\frac{d}{dt}\left(\dot{\mathbf{r}}_p \cdot \dot{\mathbf{r}}_{osc}\right). \quad (13)$$

For an impulse external force the last ("extra") terms in Eqs. (12), (13) do not disappear. But for a *quasi-periodical* particle motion we may expect, at least on average, our extra term contribution $\dfrac{d}{dt}(\dot{\mathbf{r}}_p \cdot \dot{\mathbf{r}}_{osc})$ to vanish in the finite-difference energy balance [2]. (The analogy of (13) with (12) for a quasi-periodical motion is the best for a resonance frequency oscillator with its growing average energy $\overline{E}_{osc}$.) Hence, the energy conservation law has been "reestablished" with (9) in the same way as it was done in CED.

### 3.2. Discussion of equation system (10)

Let us now see whether we really achieved what we wanted while reestablishing the energy conservation law. The system (10) can be cast in the following form:

$$\begin{cases} \tilde{M}_p \ddot{\mathbf{r}}_p = \mathbf{F}_{ext}(\mathbf{r}_p \mid t) + \alpha M_{osc}\ddot{\mathbf{r}}_{osc}, \\ \tilde{M}_{osc}\ddot{\mathbf{r}}_{osc} + k\,\mathbf{r}_{osc} = \alpha\dfrac{M_{osc}}{\tilde{M}_p}\mathbf{F}_{ext}(\mathbf{r}_p \mid t), \end{cases} \quad (14)$$

where

$$\begin{cases} \tilde{M}_p = M_p\left(1 + \eta\alpha\dfrac{M_{osc}}{M_p}\right) = M_p + \eta\alpha M_{osc}, \\ \tilde{M}_{osc} = M_{osc}\left(1 - \alpha^2\dfrac{M_{osc}}{\tilde{M}_p}\right). \end{cases} \quad (15)$$

Here in the particle equation we joined two acceleration terms in one and obtained $\tilde{M}_p$. Then we inserted $\ddot{\mathbf{r}}_p$ from the new particle equation into the oscillator one to express the pumping term via the external force, as in Eq. (8).

According to the "particle" equation in Eqs. (14), the particle inertial properties have changed: the kinetic energy of free motion acquired an addendum and with a given external force the particle solution $\mathbf{r}_p(t)$ will manifest now a different behavior due to the potential self-action contribution $\propto \eta$. For example, in case of a low-frequency external force when the "radiation reaction" is really negligible ($\omega_{ext} \ll \omega$), the equation solution corresponds to a motion of a modified mass $\tilde{M}_p$ in the external field. We did not want it. Moreover, if we weigh our particle with a spring scale in a static experiment $(\mathbf{F}_{ext})_z = M_p g - K \cdot z = 0$, we naturally obtain $M_p$ as the particle mass. Gravity force term, if present in $\mathbf{F}_{ext}$, does not acquire any "mass correction" due to our "coupling" (11), so the "bare" mass $M_p$ (if we decide to call it "bare" now) and the "correction" $\eta \alpha M_{osc}$ are in principle experimentally distinguishable. Here the situation is quite similar to that in CED despite it is often erroneously said that $m_q$ and $\delta m_q$ come always in sum and are "indistinguishable".

Our new oscillator equation in Eqs. (14) has changed too: the oscillator kinetic term and the coupling constant $\alpha$ have also acquired additional factors. It means that even in absence of external force the oscillator *proper* frequency will be different now. Indeed, with dividing (14) by $M_{osc}$, one can obtain:

$$\begin{cases} \ddot{\mathbf{r}}_{osc} + \tilde{\omega}^2 \mathbf{r}_{osc} = \tilde{\alpha} \dfrac{\mathbf{F}_{ext}}{M_p}, \quad \tilde{\omega} = \omega\left(1 - \alpha^2 \dfrac{M_{osc}}{\tilde{M}_p}\right)^{-1/2}, \\ \tilde{\alpha} = \alpha\left[\left(1 + \eta\alpha \dfrac{M_{osc}}{M_p}\right)\left(1 - \alpha^2 \dfrac{M_{osc}}{\tilde{M}_p}\right)\right]^{-1}. \end{cases} \quad (16)$$

In particular, the oscillator mass modification is similar to a photon mass (solution frequency) modification in a gauge non-invariant regularization scheme in QED. If solutions of Eqs. (14) are expanded in powers of $\alpha$, the corresponding undesirable corrections will appear in the perturbative series, as in QED.

But why all this has happened? We did not plan it in our theory development! Well, it was not immediately visible in the perturbative treatment of (10) mentioned briefly above, but our replacing a known time-dependent function $\ddot{\mathbf{r}}_p(t) = \mathbf{F}_{ext}(\mathbf{r}_p(t))/M_p$ in the right-hand side of (6) with unknown (searched) variable $\ddot{\mathbf{r}}_p$, which is in turn strongly coupled to unknown $\ddot{\mathbf{r}}_{osc}$ in (10), *was an intervention* into our oscillator equation. In other words, the oscillator equation appearance (form) in (10), i.e., its similarity to (6), is deceptive and misleading. Not having noticed this fact was our elementary mathematical error and thus we have changed the oscillator equation contrary to our intention. Therefore, there is no here any "fundamental physics of bare particles" whose "interactions" (9) modify the fundamental constants. Indeed, if in our theory development we had proceeded from the numerically equivalent wave Eq. (8), expressed via the driving force $\mathbf{F}_{ext}(\mathbf{r}_p)/M_p$ (i.e., via *another combination of variables*) instead of variable $\ddot{\mathbf{r}}_p$ in the right-hand side, we would not have spoiled the wave equation coefficients with injecting the right term $\alpha M_{osc} \ddot{\mathbf{r}}_{osc}$ into the mechanical equation. In other words, in developing a better equation system, we should have kept the right physical mechanism (preserving the right "spirit") rather than to keep its "form" (6) (or appearance) at any expense.

Now, what is the use of a formal "energy conservation law" (11) if the new equations describe some "physical systems" quite different from the original ones? We cannot keep to our ansatz (guess) (9) just on the pretext that it provides some formal "conservation law". We should find another approach to our problem – introducing a radiation reaction term.

Such was the true situation encountered first in CED and later on in QED where the equation coupling is made also with the famous self-action ansatz advanced to reestablish the conservation laws. Indeed, when the external field $A_{ext}$ is known, as in Eqs. (2), (4), the charge "interaction Lagrangian" density $(\mathcal{L}_{int})_q \propto j \cdot A_{ext}$ does not lead to troublesome particle equations. As well, when the charge motion $j_{ext}$ is known, the field "interaction Lagrangian" density $(\mathcal{L}_{int})_A \propto j_{ext} \cdot A$ doest not lead to troublesome field solutions either – the retarded potentials are good and the Maxwell equations with known $j_{ext}$ is the special relativity basis. In other words, as long as the particle and field equations are "partially decoupled", they have clear physical meaning with physical constants in them. It is the self-action ansatz $(\mathcal{L}_{int})_{qA} \propto j \cdot A$ (i.e., with both variables $j$ and $A$ considered unknown and coupled in the same form) which is responsible for spoiling particle and field equations and my purpose here was just to make it evident. The self-action ansatz preserves the equation "form", not the "spirit".

Below comes a quote how J. Schwinger perceived equation spoiling with such a coupling (see [3], page 416): "*In putting together these various equations, the physical meaning of the symbols e and m had seemed to be clear. Is it true? Not at all. Through the innocent process of combining these equations into a non-linear coupled system, the physical meanings of all the symbols have changed. They no longer refer to the physical particles…*".

Neither the principle of least action, nor the Lorentz invariance and the gauge nature of $(\mathcal{L}_{int})_{qA} \propto j \cdot A$, nor the formal Noether theorem has helped guess correct CED and QED equations, unfortunately. These equations, as we know, need further modifications like renormalization (discarding certain terms), soft diagram summation, etc., to arrive at physically meaningful results. The formal analogy with "partially decoupled" cases mentioned above did not work here – it produced an undesirable "self-amplification", figuratively speaking.

In CED and QED there was a period of searching for better theory formulations (H. Lorentz, H. Poincaré, M. Born, L. Infeld, P. Dirac, R. Feynman, A. Wheeler, F. Bopp, F. Rorhlich to name a few, see [4], [5]). P. Dirac was explicitly calling our interaction $(\mathcal{L}_{int})_{qA} \propto j \cdot A$ and the concepts behind it wrong. However, no satisfactory equations were proposed because of lack of right physical ideas. And in addition, renormalization of the fundamental constants in perturbative solutions happened to lead to nice results in some rare, but important cases; that is why constructing *renormalizable* theories has gradually become the mainstream activity. Nowadays the fact of coefficient modifications due to "interaction" similar to (9) is used in QFT not as evidence of the interaction term being wrong, but as a "proof" of the original constants (and particles) being non physical, "bare" ones [6]. The latter can be called a "theoretical discovery of non observable bare particles and their physics" and is nothing else but self-fooling. No bare particles were in our experiments, neither in the project of our theory development. Some people do not see the mathematical error made and perceive an accidental success of renormalization as a real "physical phenomenon". The key point in this errancy is an implicit and unsubstantiated claim that *our guess* (a bad theory) is right; that it *must* describe and describes the experiments, and for this (wrong) reason the blame is transferred from our bad corrections to the good original fundamental constants.

In order to make our construction (14) work, we too are going first to follow the renormalization prescription. Fortunately, in (14) it can be done exactly rather than perturbatively, i.e., in each order.

## 4. Theory Development II: Renormalizations in Equations (14)

Speaking specifically of our particle, its mass renormalization means calling the whole combination $\tilde{M}_p$ "the physical mass" and using for it the old numerical value from (1). This is equivalent to discarding the whole "correction" $\propto \eta$ in (15). After that we restore at least the right inertial properties of our solution $\mathbf{r}_p(t)$. We do not know yet if the resulting equation becomes good for the radiation reaction description, but we hope for it.

As well, we see that it is not enough to "repair" the system (14): the oscillator mass in (15) or proper frequency and the coupling constant in (16) are still different from those in (6).

Now we renormalize the oscillator mass $\tilde{M}_{osc}$ in (15) or coupling $\tilde{\alpha}$ in (16) with discarding the terms $\propto \alpha^2$. Thus, with only two "independent" renormalizations we can "obtain" an exactly renormalized equation system:

$$\begin{cases} M_p \ddot{\mathbf{r}}_p = \mathbf{F}_{ext}(\mathbf{r}_p \mid t) + \alpha M_{osc} \ddot{\mathbf{r}}_{osc}, \\ M_{osc} \ddot{\mathbf{r}}_{osc} + k \mathbf{r}_{osc} = \alpha \frac{M_{osc}}{M_p} \mathbf{F}_{ext}(\mathbf{r}_p \mid t). \end{cases} \quad (17)$$

This system only contains the physical fundamental constants in the same sense as they were contained in the original phenomenological Eqs. (1) and (6). As well, after renormalizations, we have an additional "radiation reaction" force $\alpha M_{osc} \ddot{\mathbf{r}}_{osc}$ that we have been looking for. Fortunately, this system is correct and no coefficient modification is necessary anymore. Perturbation theory in powers of $\alpha$ is still possible and it resembles that of QED with its infrared difficulties, but our perturbative solutions can be essentially improved with building new zeroth-order approximations taking into account exactly some terms depending on $\alpha$ (§ 4.2).

In terms of "interaction Lagrangian", this exact renormalization is equivalent to and can be implemented as formally adding the following "counter-terms" $L_{CT}$ to our "trial" Lagrangian $L_{Trial} = L_p^{(0)} + L_{osc}^{(0)} + L_{int}$:

$$L_{Phys} = L_{Trial} - \alpha \eta \frac{M_{osc} \dot{\mathbf{r}}_p^2}{2} + \alpha^2 \left( \frac{M_{osc}}{M_p} \right)^2 \frac{M_p \dot{\mathbf{r}}_{osc}^2}{2}. \quad (18)$$

It means subtracting from (9) the self-action contribution proportional to $\eta$ (it has never been useful in Physics, to tell the truth), leaving there the cross term $\propto \dot{\mathbf{r}}_p \dot{\mathbf{r}}_{osc}$ (partially needed for coupling Eqs. (1) and (6)), and adding a quadratic in $\alpha$ kinetic term to the oscillator kinetic energy to cancel the oscillator mass modification (oscillator "self-action") arising due to that awkward cross term. Unlike in QFT, here the subtraction can be done exactly rather than anew in each order. In particular, some part of the "new interaction" $L_{int} + L_{CT}$ can and should be taken into account exactly into new zeroth-order approximations that will give different (improved) perturbation expansions with improved description of physics. This will be shown later, in § 4.2, devoted to the perturbative treatment of Eqs. (17), and now let us discuss new physics contained in the exactly renormalized equations.

**4.1. Discussion of renormalized system (17)**

Now, let us analyze our equations (17) and their new physics, if any. First of all, the oscillator equation in it is rather "decoupled" from the particle one – it is directly influenced with the external force, as in (8). The total decoupling occurs in case of a uniform external force

$\mathbf{F}_{ext}(t)$. So, after renormalizations, we returned to the right equation and solutions for $\mathbf{r}_{osc}$, fortunately.

But let us look at the particle equation: $\mathbf{r}_p$ is now influenced with the oscillator motion in a "one-way" way! If an external force of a limited duration $T$ pushes our particle and excites the oscillator, the latter oscillates freely afterwards, but the particle gets these free oscillations as an external known force now: $\ddot{\mathbf{r}}_p \propto \ddot{\mathbf{r}}_{osc}(t)$. Did we expect such a "feedback" from the oscillator in the beginning of our program of reestablishing the energy conservation law? Didn't we expect decoupling equations in absence of external forces (acceleration)? Meanwhile the force $\propto \ddot{\mathbf{r}}_{osc}$ is now always present in the particle equation. Remember, we still do not know the particle composition and think it is a point-like object.

What should we think of our system (17)? Wrong again? Here a more fine comparison with experimental data can give us an ultimate answer since renormalizations (subtractions) do not guarantee correctness of the "interaction remainder". And let us suppose that our experimentalists discover, with sophisticated optical and acoustical measurements, that indeed, during and after force acting, our probe body actually vibrates *as a whole* (rather than changes its shape) in a qualitative agreement with the "vibrating" solution $\mathbf{r}_p(t)$ from (17), so the mechanical part of (17) is also right. We do not need to repair any equation anymore, fortunately. We are certainly lucky and now it is our understanding that needs a repair.

Indeed, we wanted better-coupled equations, namely, a "feedback" from the oscillator to the particle motion and we got it. It lasts longer than foreseen, and now the particle acceleration $\ddot{\mathbf{r}}_p$ does not influence the oscillator when the force ceased acting ($\mathbf{F}_{ext}=0$). How can that be? It can be so only if our particle belongs to the oscillator and conversely. The particle oscillations $\ddot{\mathbf{r}}_p \propto \ddot{\mathbf{r}}_{osc}(t)$ do not serve as a "pumping term" to the oscillator equation if the particle is just an oscillator piece. This is the right understanding of physics contained directly in (17). To see it better, let us introduce another dynamical variable (we join kinetic terms in (17)):

$$\mathbf{R} = \mathbf{r}_p - \alpha \frac{M_{osc}}{M_p}\mathbf{r}_{osc}. \qquad (19)$$

Then we obtain the equations:

$$\begin{cases} M_p \ddot{\mathbf{R}} = \mathbf{F}_{ext}(\mathbf{R}+\alpha\frac{M_{osc}}{M_p}\mathbf{r}_{osc} \mid t), \\ M_{osc}\ddot{\mathbf{r}}_{osc} + k\,\mathbf{r}_{osc} = \alpha\frac{M_{osc}}{M_p}\mathbf{F}_{ext}(\mathbf{R}+\alpha\frac{M_{osc}}{M_p}\mathbf{r}_{osc} \mid t). \end{cases} \qquad (20)$$

If the external force is uniform (no space arguments), the variable $\mathbf{R}$ does not have those oscillations even though the force is on. Also, after the force ceased acting, the equation for $\mathbf{R}$ describes a free motion. It looks as an equation for the center of mass (CM) of a compound system where our particle is bound to something else with an elastic potential. ($\mathbf{R}$ may be called a "smooth" variable.) The oscillator equation can be understood now as an equation for a relative/internal motion in a compound system. Hitting our particle excites the internal motion in the system and transfers some kinetic energy to the system as a whole. This is what the correct Eqs. (20) say. It may only happen if the body is a coupled (compound) system containing constituents and the external force only acts on *one* of its constituents. A feasible model for such a system ("rigid shell and a core") is already given in Fig. 1, but now we inferred it exclusively from the physical analysis of correct equations (20).

Then $M_p$ is in fact the *total mass* of the system: $M_p = M_{tot}$, and $M_{osc}$ is the reduced mass $\mu$ involved in the relative motion equation. Indeed, if we take a couple of particles (constituents) with $m_1$ and $m_2$ coupled with a spring $k$ and separate the corresponding variables $\mathbf{R} = (m_1\mathbf{r}_1 + m_2\mathbf{r}_2)/(m_1+m_1)$ and $\mathbf{r}_r = \mathbf{r}_1 - \mathbf{r}_2$, we will obtain precisely equations (20) with $M_p = M_{tot} = m_1 + m_2$, $M_{osc} = \mu = \frac{m_1 m_2}{M_{tot}}$, and $\alpha = \frac{M_{tot}}{m_2}$. The true oscillator coordinate $\mathbf{r}_r = \mathbf{r}_1 - \mathbf{r}_2$ describes, as we see now, the relative/internal motion in this compound system, and the oscillating part of $\mathbf{r}_1$ and thus the sound amplitude $A_{sound}$ are naturally proportional to it: $\mathbf{r}_1 = \mathbf{r}_p = \varepsilon \cdot \mathbf{r}_r + \mathbf{R}$, $\varepsilon = m_2/M_{tot}$, $\mathbf{R}$ being a much smoother function.

Thus, we figured out the right physics of coupling from the correct equations. It is very different from the "bare particle physics". Our constituent particles are not bare, but permanently coupled physical ones. The energy conservation law for such a system is simple and it reads: the external force work $-\Delta V_{ext}$ done on displacing the *constituent* particle from $\mathbf{r}_p(t_1)$ to $\mathbf{r}_p(t_2)$ is spent on changing the center of mass kinetic energy and on changing the relative motion (internal) energy of the compound system, both works being additive:

$$\begin{cases} -\Delta V_{ext} = \Delta\left(\frac{M_p \dot{\mathbf{R}}^2}{2}\right) + \Delta E_{osc}, \\ E = \frac{M_p \dot{\mathbf{R}}^2}{2} + V_{ext}\left(\mathbf{R}+\alpha\frac{M_{osc}}{M_p}\mathbf{r}_{osc}\right) + \frac{M_{osc}\dot{\mathbf{r}}_{osc}^2}{2} + k\frac{\mathbf{r}_{osc}^2}{2}. \end{cases} \qquad (21)$$

Lagrangian (18) in these variables is the following:

$$L_{Phys} = \frac{M_p \dot{\mathbf{R}}^2}{2} - V_{ext}(\mathbf{R}+\varepsilon\cdot\mathbf{r}_{osc}) + \frac{M_{osc}\dot{\mathbf{r}}_{osc}^2}{2} - k\frac{\mathbf{r}_{osc}^2}{2}. \qquad (22a)$$

It differs from $E$ (21) by the minus signs at the potential energies. Equation coupling occurs here via the external potential argument $\mathbf{r}_p$, not via a product like in (9).

Hence, our original problem of reestablishing conservation laws has a *reasonable physical resolution* different from our first attempt (9-10) furnished with obligatory renormalizations (subtractions). As J. Schwinger wrote in [3], page 420: *"This way of putting the matter can hardly fail to raise the question whether we have to proceed in this tortuous manner of introducing physically extraneous hypotheses only to delete these at the end in order to get physically meaningful results. Clearly, there would be a great improvement, both conceptually and computationally, if we could identify and remove the speculative hypotheses that are implicit in the unrenormalized equations, thereby working much more at the phenomenological level. ...I continue to hope that it has great appeal to the true physicist (Where are you?)."*

Let us note that system (17) is factually written in so called mixed (not yet separated) independent variables: an individual coordinate $\mathbf{r}_1 = \mathbf{r}_p$ and a relative one $\mathbf{r}_r = \mathbf{r}_1 - \mathbf{r}_2 = \mathbf{r}_{osc}$. Such a formulation contains a cross term in Lagrangian, like in Eq. (7a), and the total mass $M_{tot}$ at the constituent particle-1 acceleration $\ddot{\mathbf{r}}_1$:

$$L = \frac{m_1 \dot{\mathbf{r}}_1^2}{2} - V_{ext}(\mathbf{r}_1) + \frac{m_2 \dot{\mathbf{r}}_r^2}{2} - k \frac{\mathbf{r}_r^2}{2} - m_2 \left( \dot{\mathbf{r}}_1 \dot{\mathbf{r}}_r - \frac{\dot{\mathbf{r}}_1^2}{2} \right) \quad (22b)$$

$$\begin{cases} m_1 \ddot{\mathbf{r}}_1 = \mathbf{F}_{ext}(\mathbf{r}_1 \mid t) - m_2 \left( \ddot{\mathbf{r}}_1 - \ddot{\mathbf{r}}_r \right), \\ m_2 \ddot{\mathbf{r}}_r + k \mathbf{r}_r = m_2 \ddot{\mathbf{r}}_1. \end{cases} \quad (23)$$

This formulation and its solutions are similar *in form* to our wrong Eqs. (10) with Lagrangian (9). The exact formulation in terms of mixed variables is, of course, correct, but the perturbation theory here starts from the "wrong" masses $m_1$ and $m_2$, and a part of perturbative corrections serve here to build $M_{tot}$ and $\mu$ from $m_1$ and $m_2$ involved in the perturbative solutions. The perturbation theory for the wrong Lagrangian (9) starts, on the contrary, from the right masses $M_{tot}$ and $\mu$; that is why all corrections to masses in (10), (14) are not necessary and harmful, and after discarding them we luckily "restore" the right solutions.

System (20) can be cast in the following interesting form:

$$\begin{cases} M_p \ddot{\mathbf{R}} = \mathbf{F}_{ext}(\mathbf{R} + \alpha \frac{M_{osc}}{M_p} \mathbf{r}_{osc} \mid t), \\ M_{osc} \ddot{\mathbf{r}}_{osc} + k \mathbf{r}_{osc} = \alpha M_{osc} \ddot{\mathbf{R}}. \end{cases} \quad (24)$$

It resembles (1) with (6), but the equation for $\mathbf{R}$ is kind of "non local" – the force argument is shifted by the other variable and it provides the "radiation reaction" effect on $\mathbf{R}$ (compare it with the "local" Eqs. (17)). Note, Eqs. (24) in the zeroth order on the force gradient (or partially averaged) have the same look and numerical solutions as our original Eqs. (1) and (6). Thus, transition from the approximate Eqs. (1), (6) to the exact ones (24) may be achieved with "enlarging" the external force argument (see (19)) if these equations are understood correctly (see Section 5).

As well, equation system (24) has an advantage over the others because it is in fact a more general one and is the only appropriate formulation in situations when we do not know the studied body "composition". Indeed, in a more realistic case when the "body" cannot be disassembled into simple mechanical pieces like "points" with $m_1$, $m_2$, and a spring $k$, it is still possible to study experimentally the center of mass motion and the normal modes of the compound system in question, so an equation system à la (24) with (19) is the right phenomenological framework for that.

### 4.2. Perturbation Theory for Eqs. (20), (24)

The external force argument $\mathbf{R} + \varepsilon \mathbf{r}_{osc}$ is different from $\mathbf{R}$, so the equations in (20) or (24) are coupled in general case. It may be convenient to expand the force "around" $\mathbf{R}$ if the corresponding force difference contribution $\delta \mathbf{F}_{ext} = \mathbf{F}_{ext}(\mathbf{r}_p) - \mathbf{F}_{ext}(\mathbf{R})$ is relatively small. This "gradient term" $(\delta F_{ext})_i \approx \alpha \frac{M_{osc}}{M_p} \left( r_{osc}^{(0)} \right)_k \frac{\partial (F_{ext})_i}{\partial R_k}$ has another small perturbative parameter different from just $\alpha$; the latter is already involved in the zeroth-order solutions $\mathbf{r}_{osc}^{(0)}(\alpha)$ and

$$\mathbf{r}_p^{(0)}(\alpha) = \mathbf{R}^{(0)} + \alpha \frac{M_{osc}}{M_p} \cdot \mathbf{r}_{osc}^{(0)}(\alpha). \quad (25)$$

In terms of Lagrangian, the corresponding approximate potential energy is the following (compare it with Eqs. (3) and (7b)):

$$V_{ext}(\mathbf{r}_p) \approx V_{ext}(\mathbf{R}) - \alpha \frac{M_{osc}}{M_p} \mathbf{r}_{osc} \cdot \mathbf{F}_{ext}\left( \mathbf{R}^{(0)}(t) \right). \quad (26)$$

The advantage of this perturbation scheme is especially clear in case of a uniform external force $\delta \mathbf{F}_{ext} \equiv 0$ when the variables $\mathbf{R}$ and $\mathbf{r}_{osc}$ are completely separated and Eq. (25) turns into the exact solution.

No conceptual and/or mathematical difficulties can be expected on this way since the exact Eqs. (20), (24) are physical, have physical solutions, and the zeroth-order equations with $\mathbf{F}_{ext}\left( \mathbf{R}^{(0)} \right)$ in each of them capture already well the exact solution properties. In particular, both in Classical and Quantum Mechanical treatments the oscillator modes (including "soft" ones $\mathbf{r}_{osc} \propto 1/\omega$) are already automatically excited in this approximation and an inclusive (average) picture becomes natural. The latter result is achieved in QED only with a forced and heavy summation of divergent soft contributions to all perturbative orders because nothing from its interaction Lagrangian "patched" with counter-terms $L_{int} + L_{CT}$ is included into the zeroth-order approximation like (25) ($\alpha$ still being an expansion parameter there).

Adding an interface-air interaction makes it possible to damp the excited oscillations. Measuring the total sound energy gives the necessary data for determining $\mu$ and $\alpha$ of the true oscillator as well as the damping constant.

Interaction with the body in a direct contact (short-range force) is similar to the interaction with the total field of a charge including its short-range "near field" in CED.

With the correct physics description (19), (20), (24), we may now calculate the results of collision of any such compound bodies, Doppler and retardation effects at any distance $S$ without conceptual and mathematical difficulties.

QFT equations made physically analogous to (24) with (19) will hopefully describe the occupation number evolutions without renormalization and infrared divergence. Indeed, the QFT equations like a particle in a known external field and a field due to a known source have often reasonable physical solutions and in this sense Eqs. (20), (24) are exemplary to follow.

## 5. General Discussion

Passing through our "theory development stages" I and II, we were in fact discovering complexity of our material body. We sincerely thought the body was point-like (we applied an equation for one point!) and generally detached from the "wave" system. Experience with our toy problem teaches us we were wrong. The external force acts in fact on one of constituents of a compound system and the latter has "internal" degrees of freedom. Much more "point-like" is the center of mass instead. Correct Eqs. (20), (24) are quite comprehensible and familiar to us. They are quasi-particle equations of a compound system describing the global (CM) and the relative (internal) collective motions. If the external force is uniform, they even coincide in form with the original ones (1) and (6), (8). In case of a uniform force the energy conservation law *already holds* (no need to reestablish it) and it is so just because of different physical meaning of (separated) variables $\mathbf{R}$ and $\mathbf{r}_{osc}$. We should not couple these equations at all and we could have even deduced Eqs. (24) directly from (1) and (6) if we had initially admitted the right physical idea about our body being compound. This is what can be called a reformulation approach leading to the same physical results directly. Indeed, our renormalized equations, especially in form (24), are obviously equivalent to a theory formulated from a different physical concept – an idea whose necessity was so persistently promoted by P. Dirac [3]. In our toy model these right physical ideas are a compound character of the probe body, belonging the constituent particle-1 (point of the force application) to the "wave system", and oscillator being an "internal degree of freedom" of this compound system. Such ideas would prevent us from advancing wrong Eqs. (10), (14) with subsequent renormalizations of coefficients in them. The hint can be found already in Eq. (8) – the external force acting on a particle, acts directly on the oscillator.

We treated our probe body as simple, point-like, for two main reasons: an experimental and a human one. We humans tend to deal with as simple things as possible. And experimentally, even if we monitor the true particle-1 (really oscillating or "fluctuating") coordinate $\mathbf{r}_p(t)$, our experimental results may give us the center of mass (smooth) coordinate due to certain averaging. In our case the permanent coupling of $\mathbf{r}_p$ and $\mathbf{r}_{osc}$ in the relationship $\mathbf{r}_p = \mathbf{R} + \varepsilon \cdot \mathbf{r}_{osc}$ must have been "lost", for example, due to time averaging: $\langle \mathbf{r}_p \rangle \approx \langle \mathbf{R} \rangle$. Thus, we observed mostly an average quasi-particle coordinate $\langle \mathbf{R} \rangle$ and we thought it was a particle's, microscopic one. So one of the roots of our physical error in the theory development (9) was in our misunderstanding the meaning of phenomenological Eq. (1). Such an equation in Physics is *always* established for $\langle \mathbf{r}_p \rangle$. We should have realistically thought of (1) as of equation for an average (inclusive) quantity $\langle \mathbf{r}_p \rangle \approx \langle \mathbf{R} \rangle$.

Average (inclusive) character of some experimental notions is important for their practical observability, certainty, and determinism, but we idealize these notions and forget that they are built in reality due to summation (making correlations in an inclusive picture) and they do not exist as certain independently of them. In addition to this, the permanent coupling outlined above and factually present in the empirical wave equation (6) may be written in a way not revealing the right physics – the pumping term $\propto \ddot{\mathbf{r}}_p$ rather than $\propto \mathbf{F}_{ext}$; thus, the coupling is not considered as permanent (remember the adiabatic hypothesis in QFT) although it is such. Then attempts to couple coupled already things fail and we invent "renormalizations" and other weird "physics" on the go to get out of our conceptual impasse.

I did not write the damping explicitly, but I meant it: the sound observations leading to Eq. (6) are only possible due to oscillator's gradually transmitting its energy to the air or directly to a measuring device and this constitutes the role of the environment including an "observer". Even after damping out oscillations, our steady particle-1 remains permanently coupled within the oscillator and ready for new adventures. In Quantum Mechanics such a charge permanently coupled within the electromagnetic field oscillators (called an "electronium") is described with elastic and inelastic state-dependent form-factors briefly outlined in [7]. In 1948 T. Welton even proposed something analogous to (17), (22), and (24): his electron was permanently influenced with an "external force of electromagnetic field oscillators" (something like $\mathbf{r}_{osc}^{(0)}(t)$ in Eq. (17)) and that led to the main part of the Lamb shift [8]. However, his estimations were considered qualitative, probably because such a "one-way" influence was hard to imagine (oscillators exist "everywhere in space" and influence the electron, but not vice versa). Had he figured out that the electron belonged to the field oscillators and that the latter described the relative, collective motions in a compound system (quasi-particles), the QED development might have taken another route.

Above we arrived at the right equation system after fulfilling an exact renormalization of two coefficients. Without profound analysis it may give an impression that renormalization is a good way of doing physics as it works. But let us not fool ourselves. Renormalizations may sometimes work because the correct equations (perturbative solutions in QFT) may be so simple that they can be guessed right from the obviously wrong ones. We constructed coupled Eqs. (10) from the original ones (1), (6), where the original ones worked nearly fine: the fundamental constants are defined precisely from them. To couple better (1) and (6) we first introduced wrong interaction terms that couple equations indeed, but such a trial coupling modified masses (equation coefficients), and then we decided to discard these obviously harmful modifications. Both our actions (1), (6) → (10) and (14) → (17) nearly canceled each other – the unexpected and unnecessary mass "corrections" were removed by hand (18) for the new equations to be compliant with the old ones (1) and (6). There was only a little chance that the net remainder of these zigzag "development stages", $\alpha M_{osc} \ddot{\mathbf{r}}_{osc}$, would be good and Eqs. (1), (6) would become coupled correctly in the end. CED with its remainder $\propto \dddot{\mathbf{r}}_q$ and runaway exact solutions, as well as all non renormalizable QFT, are bright examples of a failure of such a "self-interaction approach". Thus, a "considerable success of renormalization" is a fluke [9].

Since it is *we* who changed the coefficients twice, it is useless to study relationships between "bare" constants and "physical" ones: there were no bare ones in (1) and (6), but quasi-particle parameters $M_{tot}$, $\mu$, $\omega$, etc. The notion of a "bare" constant with a "wild value" was invented by people when they tried to keep bad (wild) corrections as legitimate ones and at the same time to not contradict the experimental measurements in accidentally renormalizable theories; thus the original constants were made "guilty": wild and cut-of dependent.

In reality it is the coefficient corrections $\delta m$, $\delta e$, etc., due to the wrong trial "interaction Lagrangian" who are bad, not the original constants like ours $M_p$, $M_{osc}$, $\omega$, and $\alpha$, and it is precisely these bad corrections who are finally *entirely* discarded.

Hence, a "bare particle physics" is not a real physics at all, but a weird picture and a wrong interpretation imposed when people postulate wrong theoretical constructions as right and "force" them to describe the reality. Only a strong belief in correctness and uniqueness of the wrongly coupled equations and an *accidental* "success" of renormalizations make some accept this "bare particle physics" [6]. Indeed, the very first effect of coupling to something in quantum mechanics is quantum mechanical smearing and energy level formation rather than a "bare vacuum polarization" around a "steady bare charge", as if it were in classical dielectrics.

As well, the absence of soft radiation in the first Born approximation in QED (i.e., existing elastic processes, whose probability is zero in Nature, instead of inclusive ones, whose probability is unity) is a crying sign of a bad initial approximation apparently caused with a too superficial understanding of how the correct coupling must in reality be done and treated (see (19), (24) and (25) in § 4.2). Thus, the mainstream renormalization ideology is based on wrong ideas and notions, and even good agreement with experimental results (i.e., good physics at hand, like in Eq. (8)) does not help to infer the right conclusions because of imposed belief into "bare particle physics".

## 6. Conclusions

We have seen that experimentally established Eqs. (1) and (6) could be misunderstood and coupled in a wrong way like in our simple mechanical case of "ringing a bell" due to our physical and mathematical errors. First, we advanced a wrong interaction Lagrangian (9) by analogy with (7a) that looked natural and innocent; next, we modified the obviously wrongly coupled equations with renormalizing masses in their kinetic terms. In our toy model the renormalizations could be fulfilled exactly in the equations or in the total Lagrangian. Fortunately the right microscopic equations were so simple that the renormalizations were practically the only "repair" to obtain them. Renormalized equations turned out to be the right ones and problem-free; however they revealed a different and surprising physics – that of permanently coupled constituents rather than a "bare particle physics". Thus, renormalization is a transition from "self-action" to a permanent interaction of constituents of a compound interacting system. This is how we then figured out that our phenomenological "mechanical" and "wave" Eqs. (1) and (6) corresponded factually to the exact Eqs. (24) written in terms of separated variables (center of mass and relative motion variables), that is why they should have been coupled differently. Finally, we understood that the observed masses and frequencies corresponded to quasi-particles (or normal modes) of our compound (permanently interacting) system. Quasi-particles are a familiar and a natural language for interacting systems and it is favourably different from unobservable "bare particles". In this respect our toy model is quite instructive. It means our original phenomenological equations could have been coupled in a different, correct way immediately and directly if we had initially had the right physical ideas about the observed phenomenon: what is permanently coupled in Nature, should be implemented so in our theory. The mechanical model from Fig. 1 demonstrates it eloquently. A better initial approximation leads to a better perturbative series – the latter becomes a series of numerically small terms. Popular references to our not knowing physics of short distances to justify renormalization in wrongly coupled equations are not serious. Any theory is incomplete, it is true, but it is not bound to produce catastrophes. In a correct quantum mechanical formulation of an incomplete theory (à la (24), for example) the high energy modes (wrong or right) are not physically excited and their technical (perturbative) contributions are normally reduced to elastic form-factors close to unity which is not harmful at all.

This whole story is quite similar to QED except for in QED the right understanding of coupling has not been reached yet. Indeed, originally in QED the interaction Lagrangian was meant to simply change the occupation numbers of known particles, but it gave corrections to the equation coefficients too. The latter drawback was repaired with a coefficient renormalization which is generally equivalent to a theory reformulation, but this fact is no longer underlined. The accidental "success of renormalization" in QFT should have been understood as a strong invitation to revise our wrong equations leading to wrong "bare physics" for the sake of having an initially better physical formulation (which is hopefully feasible). This is what P. Dirac and many others pointed out to. Instead, the renormalization is now given a fundamental status and the permanent coupling in theory is switched off in asymptotical time regions. Thus the electron and field oscillators are still thought to become independent rather than permanently coupled.

I still hope the point of no return in this problem is not behind yet. In my toy model a *reformulated theory* is a theory "exactly renormalized" in the very beginning (i.e., non perturbatively, see Lagrangians (18) and (22)) so it contains only physical constants giving the same results, but is based on *conceptually different physics* (see the caption under P. Dirac photo in [3], p. XXIV). Generally a "reformulated theory" is not obligatorily reduced to an exactly renormalized one because renormalization is not a reliable way of doing physics.

A "quasi-particle formulation" of QFT is not so difficult to accept and we should not be embarrassed to construct it. Let us remember the boundary conditions – even they are simplified (approximate) solutions of "field" and "matter" interactions, so we in fact always deal with quasi-particles in compound interacting systems. These ideas, I believe, will help eliminate wrong and harmful notions from physics as well as will help reformulate some non renormalizable theories.

# References

[1]  Landau, L.D. and Lifshitz, E.M, *The Classical Theory of Fields*, **Vol. 2** (4th ed.), Reed Educational and Professional Publishing Ltd, 1975, Problem to §17, p. 52.
[2]  Feynman, R., Leighton, R., and Sands, M, *Feynman Lectures on Physics*, **Vol. 2**, Addison-Wesley Publishing Company, Inc. 1964, Chapter 28, p. 28-7.
[3]  Jagdish Mehra (editor), *The Physicist's Conception of Nature*, D. Reidel Publishing Company, Dordrecht-Holland/Boston-USA, 1973.
[4]  Rohrlich, F, *Classical Charged Particles* (3rd ed.), World Scientific Publishing, Singapore, 2006.
[5]  Laurie M. Brown (editor), *Renormalization from Lorentz to Landau (and beyond)*, Springer-Verlag, New York, 1993.
[6]  G. 't Hooft, *"Renormalization and gauge invariance"*, December 2009, Chapter 2 [Online], Available at: www.staff.science.uu.nl/~hooft101/gthpub/GtH_Yukawa_06.pdf.
[7]  Kalitvianski, V, "Atom as a Dressed Nucleus", *Cent. Eur. J. Phys.* 7(1), 1-11, March 2009.
[8]  Welton, T.A, "Some Observable Effects of the Quantum-Mechanical Fluctuations of the Electromagnetic Field", *Phys. Rev.* 74, 1157-1167, Nov.1948.
[9]  Dirac's Lectures in New Zealand, Quantum Mechanics, (1975), http://www.youtube.com/watch?v=vwYs8tTLZ24, t = 59:10 min.